# Microscopic Approach to Analyze Solar-Sail Space-Environment Effects

**Roman Ya. Kezerashvili and Gregory L. Matloff**

*Physics Department, New York City College of Technology,*
*The City University of New York*
*300 Jay Street, Brooklyn, NY 11201, USA*
*Email:* rkezerashvili@citytech.cuny.edu gmatloff@citytech.cuny.edu

## Abstract

Near-sun space-environment effects on metallic thin films solar sails as well as hollow-body sails with inflation fill gas are considered. Analysis of interaction of the solar radiation with the solar sail materials is presented. This analysis evaluates worst-case solar radiation effects during solar-radiation-pressure acceleration. The dependence of the thickness of solar sail on temperature and on wavelength of the electromagnetic spectrum of solar radiation is investigated. Physical processes of the interaction of photons, electrons, protons and α-particles with sail material atoms and nuclei, and inflation fill gas molecules are analyzed. Calculations utilized conservative assumptions with the highest values for the available cross sections for interactions of solar photons, electrons and protons with atoms, nuclei and hydrogen molecules. It is shown that for high-energy photons, electrons and protons the beryllium sail is mostly transparent. Sail material will be partially ionized by solar UV and low-energy solar electrons. For a hollow-body photon sail effects including hydrogen diffusion through the solar sail walls, and electrostatic pressure is considered. Electrostatic pressure caused by the electrically charged sail's electric field may require mitigation since sail material tensile strength decreases with elevated temperature.

## 1. Introduction

The solar sail is one of the very few proposed space-propulsion methods that may eventually be applied to interstellar exploration and travel [1,2]. For such an application, the sail must be deployed as close to the Sun as possible. A solar sail can generate a high cruise speed by using a close solar approach since the dynamic efficiency of the solar sail as a propelling device increases as the inverse square of the distance from the Sun increases. It is necessary therefore, to use sail materials that are highly reflective, low mass, and heat tolerant [3], when determining the maximum reachable near-Sun region for the solar sail spacecraft trajectory perihelion.

But that is not the entire story. The near-Sun environment is a very dynamic place. As well as a high photon flux, there is a stream of electrically charged particles that is ejected from the Sun. At close perihelion distances, this corpuscular part of solar radiation will be much greater than at Earth's solar location. This part of solar radiation is highly variable in terms of both velocity and density.

The interactions of solar-sail material with positive and negative Sun-generated ions must therefore be considered by interstellar mission planners. Solar energetic particle events can have a significant effect both on the operations and design of a solar sail spacecraft. The high near-Sun flux of ionizing solar radiation—gamma rays, *X*-rays, and ultraviolet (UV)—presents other design issues. The near-Sun space environment is very diverse including of all electromagnetic radiation frequencies, energetic charged particles, plasmas, dust, micrometeoroids and debris. The aspects of the space environment we concern ourselves with here include the electromagnetic radiation and energetic electrons, protons and α-particles and their interaction with a solar sail material.

The paper is organized as follows: In Sections 2 and 3 we discussed the sail types, near-Sun space-environment effects and structure of the Sun radiation. Requirements for a thickness of solar sail material and its dependence on the frequency of the solar radiation and temperature of a solar sail material through the electrical conductivity and the dielectric constant implicit dependence on the frequency and the temperature dependence of the conductivity are presented in Sec. 4. In Sec. 5 we investigated the interaction of the electromagnetic and corpuscular parts of solar radiation with a solar sail. A diffusion of inflation gas in hollow-body sail is presented in Sec. 6. Conclusions follow in Section 7.

## 2. Sail Types and Space-Environment Effects

Researches in sundiving solar sails have led to consideration of many sail materials and configurations. Broadly speaking, there are two basic classes of solar-sail configurations with applicability to interstellar travel. The first class includes "puller" sails, such as the parachute sail, which drags payload behind a single sheet of hyper-thin solar-sail material. The payload is attached to the sail by strong cables. The second class includes "pusher" sails such as the hollow-body or pillow sail [4, 5]. In this class, the sail is an inflatable structure containing a suitable inflation gas to maintain sail internal pressure. Unlike the parachute sail, which pulls the payload as it accelerates by solar radiation pressure, the accelerating hollow-body sail pushes against the payload. Although the requirement for a double sail thickness renders the hollow-body sail more massive than an equal-radius parachute sail, cables are not required.

As discussed in Refs. 6 and 7, both sail classes will be affected by sail-material ionization due to interaction with the solar-wind and high-energy solar-photon flux. Designers of pusher-sails must also consider the interaction between the fill gas and the walls of the solar sail, therefore a diffusion of the fill gas through the walls of the hollow-body sail.

## 3. Structure of Solar Radiation

Let us consider the components of the solar radiation that interact with the solar sail. Solar radiation has two components: the electromagnetic radiation and radiation of low- and high-energy elementary particles like electron, protons, neutrinos and ions of light nuclei emitted by the Sun. The solar radiation mainly results from solar flares, a solar wind, coronal mass ejections and solar prominences.

First we analyze the interaction of the electromagnetic radiation with the solar sail material. The photoelectric effect, the Compton effect and creation of electron-positron pairs are the main processes and they play a crucial role in the attenuation of the ultraviolet, soft and hard $X$-ray and $\gamma$-ray by a solar sail when electromagnetic radiation interacts with the solar sail atoms. When photons exceed the threshold for nucleon knockout from solar sail material nuclei, a nuclear photoeffect occurs. All photons that produce the photoelectric effect or create the electron-positron pairs give up all their energy to the sail.

The main sources of electrons and protons that interact with the solar sail are the solar wind, coronal mass ejections and solar flares. The first two produce electrons with energies from about 0.01 eV up to a few hundreds of eV and the flares are a source of high energy electrons with energies from 1 MeV up to hundreds of MeV. The energy spectrum of the solar protons extends from 0.2 keV to a few tens of keV for solar wind and coronal mass ejection protons and up to a few GeV for solar flare protons.

Thus, we can conclude that basic components of the solar radiation that will interact with the solar sail are electromagnetic radiation with energy from a few tenths of eV to hundreds of MeV, and electrons and protons in the energy spectrum from a few tenths of eV to hundreds of MeV and up to GeV.

Solar electromagnetic radiation, electrons and protons will interact with the solar sail in two manners: with the foil of the sail and with the suitable inflation gas for the hollow-body solar sail. The interaction of the solar electromagnetic radiation with the solar sail not only accelerates the spacecraft but also induces the diversity of physical processes that change the physical properties of the sail. As it is shown in Ref. 7, for high-energy photons, which are $X$ - and $\gamma$ -rays portion of the spectrum, the sail is mostly transparent. However, metallic sails, such as, for example, beryllium sails, will be partially ionized by the solar UV radiation, as is shown in Ref. 7. In other words, almost 7% of the extraterrestrial solar electromagnetic radiation will only ionize the sail producing the surface charge distribution. The visible wavelengths of the incident extraterrestrial solar electromagnetic radiation are mostly reflected by the solar sail depending on the coefficient of reflection of the solar sail material. But a part of the visible radiation as well as the infrared portion of the spectrum with wavelengths greater than 780 nm will be mostly absorbed by the solar sail causing the heating of the sail and, therefore, increase of its temperature. For the hollow-body class sails the diffusion of the inflation gas through sail's wall with an increase of temperature is the other important issue that required a consideration. Thus, the ionization, the heating of the sail and diffusion of the inflation gas are three major processes, which require mitigation and should be considered as the restrictions for the solar sail spacecraft deployment trajectories at the perihelion of an initially elliptical transfer orbit to a parabolic or hyperbolic solar orbit depending on the lightness factor of the solar sail.

## 4. Requirements for thickness of solar sail

One of the key design parameter, which determines the solar sail performance, is the solar sail areal mass, which depends on the thickness and density of the sail material as follows:

$$s = \rho d , \qquad (1)$$

where $d$ is the thickness of the sail and $\rho$ is the density of the sail material. A deployment of the sailcraft depends on the sail areal mass as well as the sail pitch angle, which is defined as the angle between the normal of the sail surface and the incident radiation. It is clear from Eq. (1) that to obtain a high performance sail we should select among the materials with the same optical properties, the material with low density and use a thin foil of this material for the solar sail. On the other hand, the ionization of a solar sail as well as the diffusion of the fill gas for a hollow-body sail also depends on the thickness of a sail material. It is obvious that when thickness of the sail increases the penetration of the solar radiation decreases and, therefore, the attenuation increases. If the solar sail film is too thin it becomes transparent for the part of the spectrum of the solar electromagnetic radiation and, therefore, this part of solar radiation is lost for the acceleration of the sailcraft. The question is how thin the foil for a solar sail should be that it would produce the acceleration of the sailcraft based on the maximum reflection of solar radiation? For the electromagnetic part of the solar radiation electromagnetic fields inside a metallic conducting foil decay rapidly with depth. The distance it takes to reduce the amplitudes of the electromagnetic field by factor of $1/e$ ($e$-folding distance) is a skin depth and it is a measure of how far electromagnetic wave penetrates into the conducting metallic foil. It is obvious that the foil thickness should be always larger than the skin depth otherwise the solar sail material will be transparent for electromagnetic radiation. Following the standard electrodynamics approach [8] by applying the system of Maxwell's equations for linear conducting media we find the minimum foil thickness that provides the maximum reflectivity and investigate dependence of this minimum thickness on wavelengths of solar radiation as well as on temperature. As it is shown in Ref. 9, the condition for the thickness of the solar sail that performs the acceleration of the sailcraft based on the maximum reflection of the solar radiation should be at least the following:

$$d = \omega^{-1}\left[\frac{\varepsilon(\omega)\mu}{2}\left(\sqrt{1+\left(\frac{\sigma(\omega,T)}{\varepsilon(\omega)\omega}\right)^2}-1\right)\right]^{-1/2}, \quad (2)$$

where $\omega$ is a frequency of solar radiation, $\varepsilon(\omega)$ and $\sigma(\omega,T)$ are the permittivity and conductivity of the metallic foil, respectively, and they are the functions of the frequency. The permeability $\mu = \mu_0 = 4\pi\times10^{-7}\,\frac{\mathrm{T\cdot m}}{\mathrm{A}}$, assumed the solar sail material to be nonmagnetic. Using Eq. (2) we can estimate the minimal thickness of the solar sail required to achieve maximum reflection of the solar light for the given optical properties of the metallic foil. The required thickness of the metallic foil is governed by three properties of the solar sail material: the permittivity $\varepsilon$, the permeability $\mu$, and the conductivity $\sigma$ of the metallic foil. Actually, each of these parameters depends to some extent on the frequency of the electromagnetic wave and the temperature. Indeed, as its follows from Eq. (2) the minimal thickness depends on the frequency - apart from the explicit dependence on the factor $1/\sqrt{\omega}$ and it also has implicit dependence, because the conductivity $\sigma(\omega,T)$ and dielectric function $\varepsilon(\omega)$ are frequency functions. Also the minimal thickness depends implicitly on the temperature through the temperature dependence of the conductivity.

The required thickness of the solar sail material is fluctuating depending on the temperature and frequency, therefore, we need to determine the minimal thickness of the sail material that would enable it to achieve the reflection of all solar spectrum frequencies at the temperatures of the near-Sun environment.

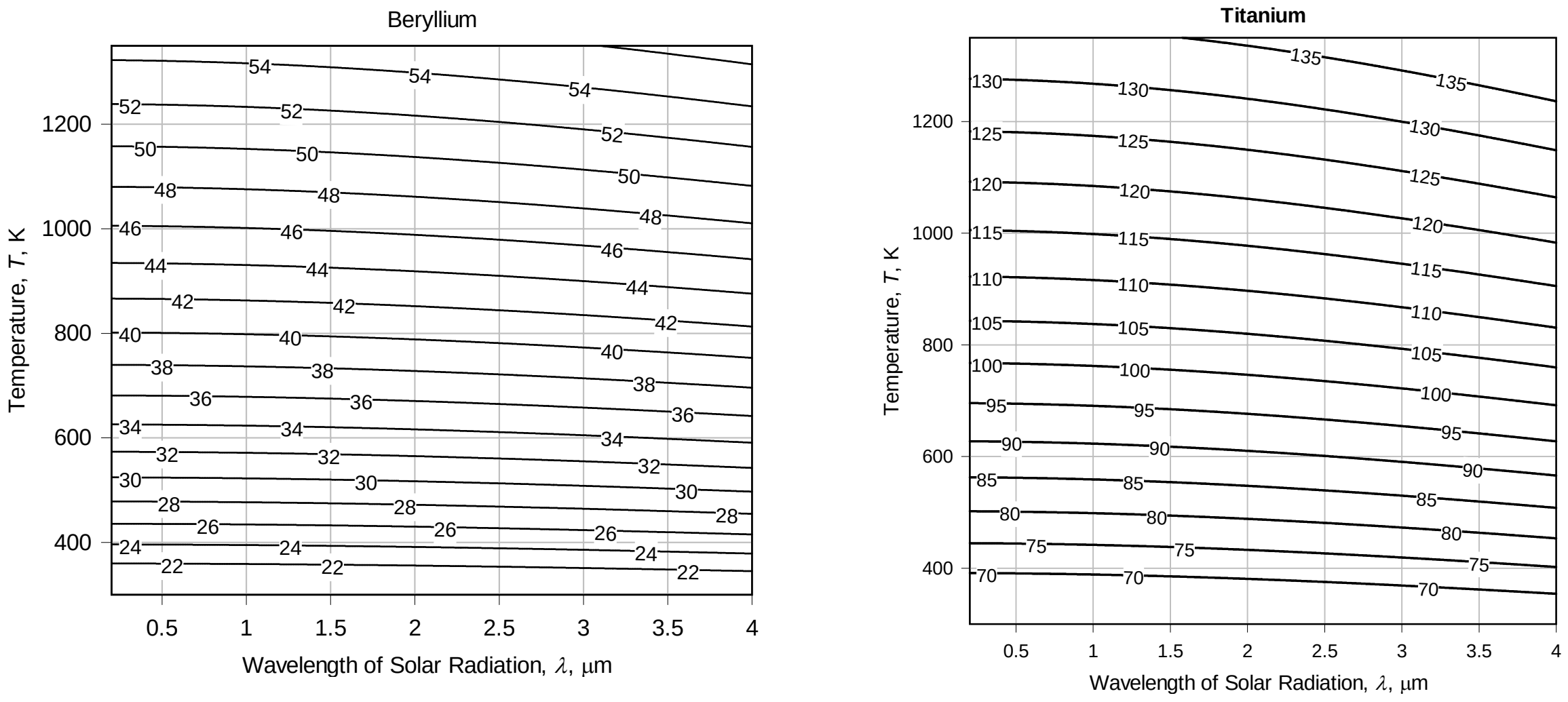

Fig. 1. The dependence of the thickness of the solar sail foil on the solar radiation wavelength and temperature for beryllium and titanium. The numbers on the curves indicate the thickness of the foil in nm.

In studying the dependence of the sail material thickness on the temperature and on the solar electromagnetic spectrum frequency (wavelength) we use Eq. (2) and have assumed the solar sail material to be nonmagnetic. The calculations were performed for beryllium and titanium. The results of the calculations are presented in Figs. 1 and 2, where the values of the required thicknesses of the solar sail material providing the best reflection of electromagnetic radiation are plotted as a function of wavelength of the solar radiation and temperature. The general behavior of these dependences is such that the increase in the temperature requires increases in the thickness of the sail foil to keep its best reflection ability. The analysis of Figs. 1 and 2 shows that the thickness of the foil exhibits the negligible dependence on the wavelength at low temperature and weakly decreases for all wavelength ranges at high temperature. We also observe much stronger dependence of the minimal thickness on the temperature, especially in the range corresponding to the visual part of the solar radiation spectrum. However, both solar sail materials exhibit strong dependence of the thickness on the wavelength in the wavelengths range $0.2\,\mu\text{m} < \lambda < 0.8\,\mu\text{m}$ and this dependence becomes stronger when temperature increases. The corresponding 3D plots for the thickness dependence on the wavelength and temperature for beryllium and titanium are presented in Fig. 2. Analysis of the results of calculations shows that the minimal thickness requirement that provides the best reflection and absorption of all solar radiation wavelengths for temperature range about 700-900 K is about 40 nm for beryllium and aluminum and about 110 nm for scandium and titanium. Using these thicknesses we can estimate the degree of ionization of the corresponding metallic foil as well as find the diffusion rate of the fill gas for a hollow body sail.

## 5. Interaction of Solar Radiation with Solar Sail

The intensity of the electromagnetic radiation penetrating a layer of material with thickness $x$ is given by the exponential attenuation law

$$I = I_0 e^{-\mu x}, \tag{3}$$

where $I_0$ is the incident photons intensity, $x$ is a penetration distance in g/cm$^2$ and $\mu$ is a mass attenuation coefficient. The mass attenuation coefficient $\mu$ is a basic quantity used in calculations of the penetration and the energy deposition by photons ($X$-ray, $\gamma$ -ray, bremsstrahlung) in materials. The mass attenuation coefficient is proportional to the total photon interaction cross section. The total cross section can be written as the sum of contributions from the principal photon interactions with a metal foil. As a result for the linear attenuation coefficient we have

$$\mu = (n\sigma_{ape} + nZ\sigma_C + n\sigma_{pair} + n\sigma_{npe})/\rho\,. \tag{4}$$

In Eq. (4) $n$ is the number of metal atoms in the unit volume, $\rho$ is a density of the solar sail material, $\sigma_{ape}$ is the atomic photoelectric effect cross section, $\sigma_C$ is the Compton scattering cross section, $\sigma_{pair}$ is the cross section for electron-positron production, and $\sigma_{npe}$ is the photonuclear cross section. The attenuation coefficient is a function of photon energy because each cross section depends on the photon energy. The first term in Eq. (4) is dominant for a low photon energy, the second term becomes important for photon energy up to hundreds of keV, the third term plays the major role for photons energy higher than a few MeV, and the last term gives the dominate contribution for high-energy photons when they are interacted with a solar material nuclei. Each cross section is a function of the photon energy. Using the corresponding cross section for the considered processes we can estimate the reduction of the radiation through the solar sail.

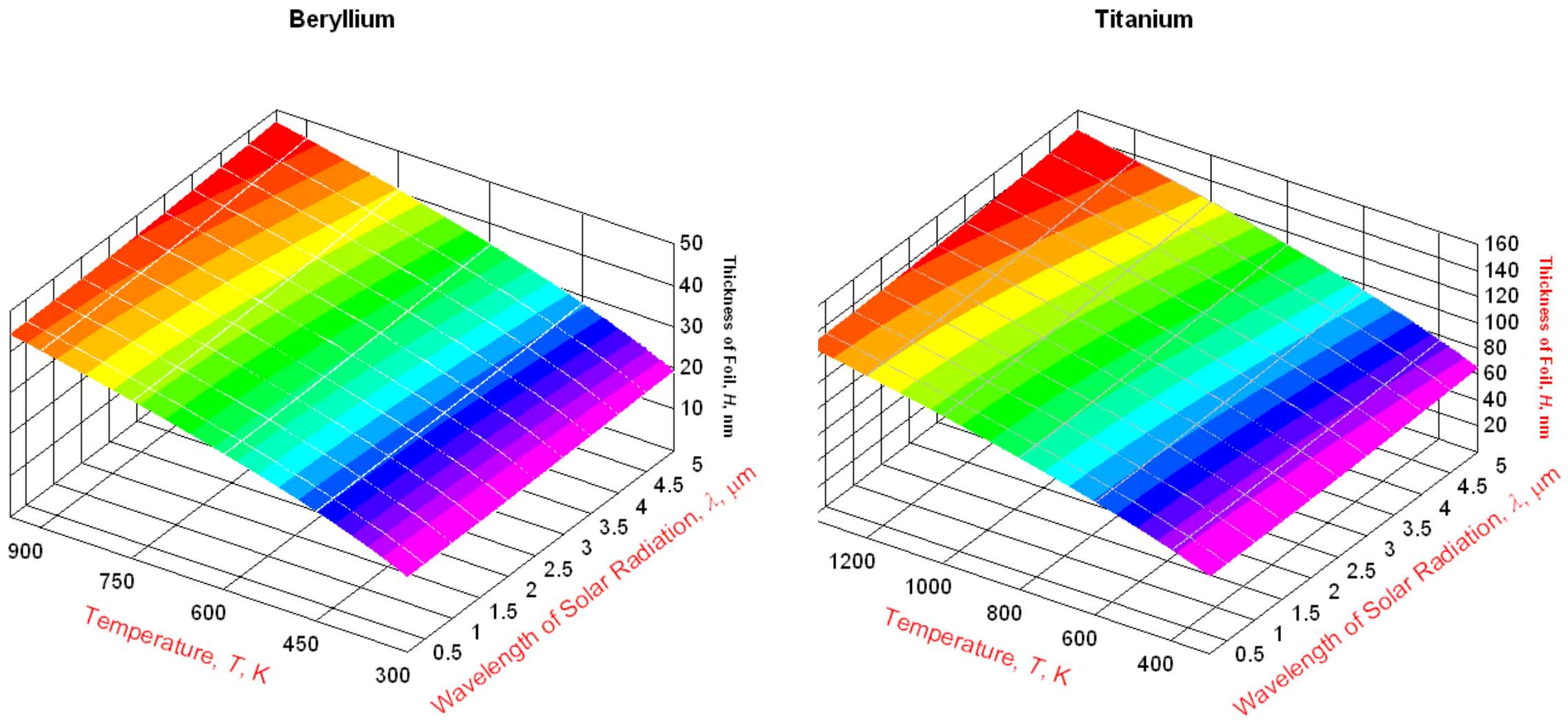

Fig. 2. The dependence of the thickness of the solar sail foil on the solar radiation wavelength and temperature for beryllium.

Electrons, protons and α-particles will also interact with the solar sail foil. As well as penetrating through the foil they interact with molecules of the inflation fill gas. At low energy, electrons scatter on the sail material atoms and basic physical processes are the excitations and ionization of the atoms of solar sail material. We can use Eq. (3) to determine the solar electron flux reduction with the coefficient of linear attenuation as

$$\mu_{electron} = (n\sigma_{ex} + nZ\sigma_i)/\rho\,, \qquad (5)$$

where $\sigma_{ex}$ is the sum of total cross sections for excitation processes in different states, and $\sigma_i$ is the sum of total cross sections for the ionization with excitation. The energy dependence of the cross section results in the energy dependence of the linear attenuation coefficient.

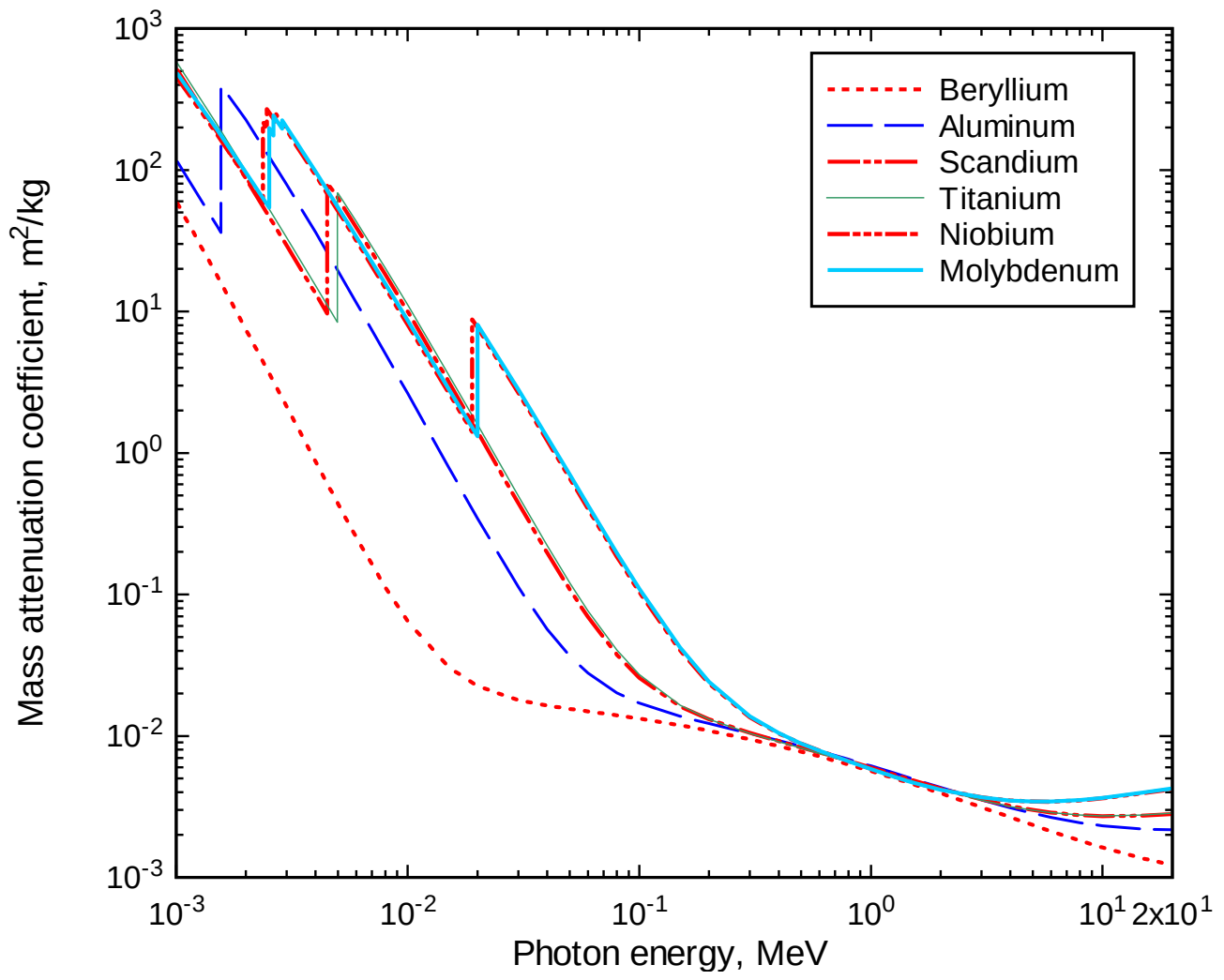

Fig. 3. The mass attenuation coefficient as a function of photon energy, for six different candidate solar-sail materials**.**

The scenario with the maximum value of the cross section for each process is conservatively assumed to better understand the processes' influence on the solar sail. To make a meaningful estimate of all effects of the interaction of photons with the solar sail material leading to its ionization, we consider the upper limits of the experimental and theoretical values of the cross sections for all considered processes.

Using the current photon interaction database of the National Institute of Standards and Technology and the tables given by Hubbell in the International Journal of Applied Radiation and Isotopes [10] we calculated the mass attenuation coefficient for different candidates for a solar sail material considered in Ref. 11. Fig. 3 presents the energy dependence of the mass attenuation coefficient for photons of energy from 1 keV to 20 MeV penetrating through foil of different metals considered as a candidate solar-sail material. For photon energies above about 4 MeV, the energy absorption is mainly due to electron-positron pair creation; below 100 keV it is due mainly to the atomic photoelectric effect; and in the intermediate range, it is due mainly to Compton scattering. The comparison of the mass attenuation coefficient in beryllium, aluminum, scandium, titanium, niobium and molybdenum shows that in beryllium it is smaller than in the other metallic foils, for scandium and titanium the coefficient almost the same. The same behavior can be observed for niobium and molybdenum. Broadly speaking, we can attribute the smaller attenuation and absorption in beryllium to the smaller density, - beryllium has fewer electrons per unit volume than aluminum, scandium, titanium and niobium and hence a photon is less likely to encounter an electron in a given thickness of material. However, the differences in details between the shapes of the curves in Fig. 3 must be attributed to the differences of the electrons arrangements in aluminum ($1s^2 2s^2 2p^6 3s^2 3p^1$), scandium ($1s^2 2s^2 2p^6 3s^2 3p^6 3d^1 4s^2$), titanium ($1s^2 2s^2 2p^6 3s^2 3p^6 3d^2 4s^2$), niobium ($1s^2 2s^2 2p^6 3s^2 3p^6 3d^{10} 4s^2 4p^6 d^4 5s^1$), and in beryllium ($1s^2 2s^2$) atoms. The scandium and titanium differ only in the number of electrons in the 3d subshell and since their properties are mainly due to their 4s electrons the energy dependence of the mass attenuation coefficients are quite similar. It is important to mention that when the energy of a photon increases from 1 keV to 10 keV the coefficients decreases by the factor $10^3$. For photon energies more than about 130 keV the mass attenuation coefficient is almost the same for all solar-sail material considered. In other words, for this energy region there is no distinction between all materials considered as a candidate for a solar sail and they attenuate photon equally.

We studied the dependence the ratio $I/I_0$ on the thickness of the solar sail foil for the different photons energies. As the example we illustrate this for beryllium foil. Fig. 4 shows the results of the calculation of the dependence of this ratio on the thickness of the beryllium foil for the different photons energies. For UV radiation (10 eV) about 55 % of the incident flux intensity will ionize the 40 nm beryllium foil and for 50 eV and 100 eV photons only 18 % and 4 % of incident photons will interact with the solar sail. For higher energy photons the beryllium solar sail will be transparent. When the foil thickness increases the degree of ionization increases also. For example, for the 50 nm thickness foil about 60% UV will ionize the sail and the role of 54 keV radiation in ionization is only about 4%. Thus, beryllium is highly transparent to *X*- and *γ*-ray, but the processing and use of beryllium is risky due to potential toxicity.

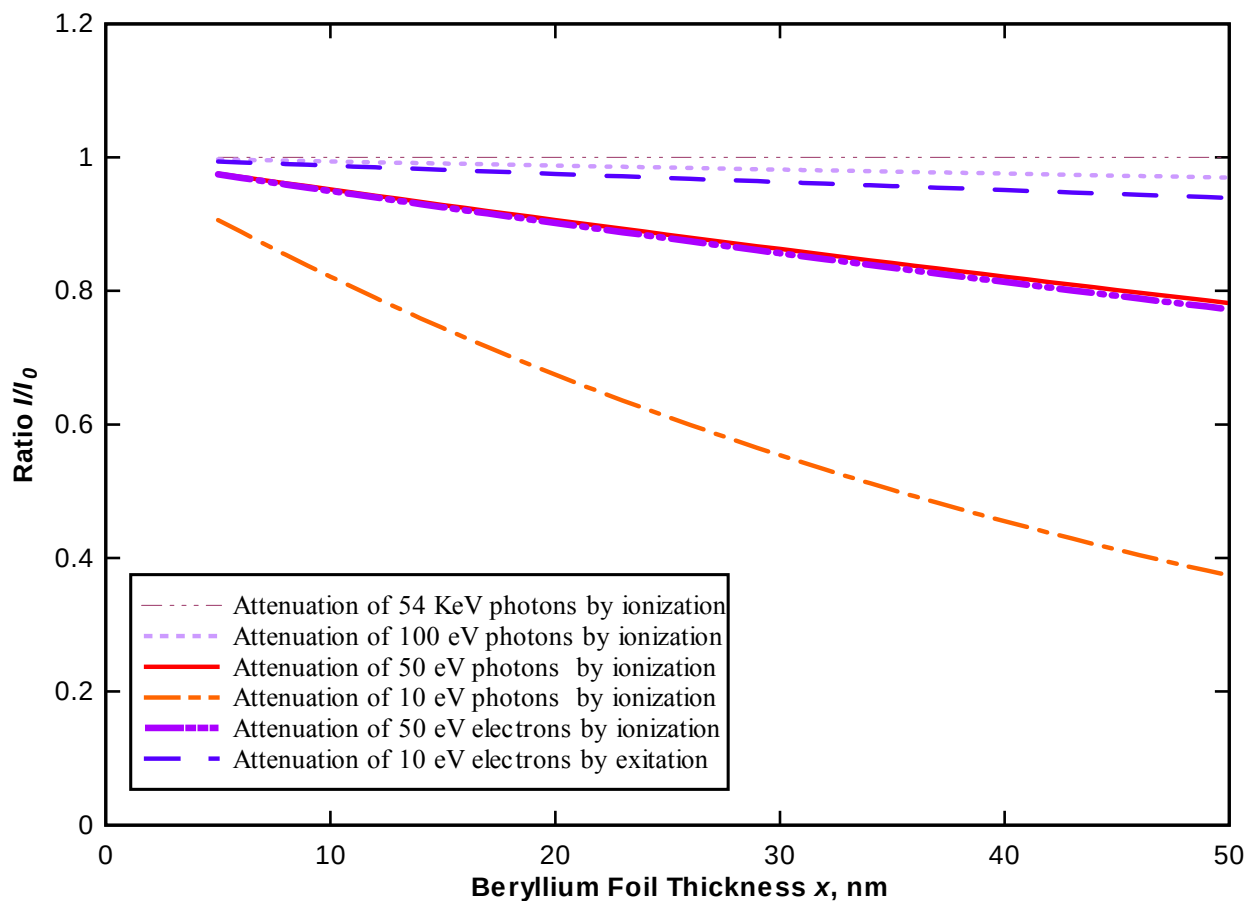

Fig. 4. Dependence of the ratio $I/I_0$ on the thickness of beryllium foil for different photon and electron energies.

For calculation of the linear attenuation coefficient for electrons related to the excitation processes we used the total cross sections for elastic and excitations scattering from Ref. [23] and for the linear attenuation coefficient related to the ionization with excitation processes we used the best available cross sections for electron impact ionization of Be atoms from Ref. [24]. Our study Ref. 6 indicate that the linear attenuation coefficient for the ionization with excitation processes is relatively smaller than for the excitation processes and has the maximum at 50 eV. For electron energies more than 200 eV the attenuation coefficients for both processes are comparable. We used available theoretical and experimental data for the total cross sections for excitation and ionization with excitation processes to find the reduction of low-energy electrons flux in the beryllium foil.

To make the most conservative prediction of the dependence of the ratio $I/I_0$ on the thickness of the beryllium foil for the excitations and ionization processes, we use the maximum value of the attenuation coefficients for the corresponding processes. Results of these calculations are presented in Fig. 4. The results indicate that the sail ionization through electrons impact is relatively much smaller than by UV radiation and reduction of the low-energy electrons proceeds through the elastic and inelastic scattering of the electrons on the beryllium foil. More than 80% to 95% of the impact electrons, depending on the foil thickness, will penetrate the sail and interact with inflation gas molecules through inelastic scattering with excitation in different states and processes of ionization in case of the hollow-body sail.

Now let us consider the interaction of protons and α-particles with the solar sail material. When electrons protons and α-particles pass through the sail material there is the energy loss for electrons, protons and α-particles. For electrons, this energy loss is the result of collisions with background atomic electrons and bremsstrahlung radiation when they are strongly deflected by the nuclei of solar sail material atoms. For protons and α-particles, it is caused by electronic and nuclear collisions. The Bethe formula [12] gives the collisional stopping power for relativistic electrons. Using the energy loss rate, the stopping range for electrons, protons and α-particles can be determined. Fig. 5 presents the stopping range for protons and α-particles of various energies for a different candidate solar-sail material**.** We use the data from the ESTAR and ASTAR database [13] and the program, which calculates the stopping range for protons and α-particles in different candidates for solar sail materials. Comparing these stopping ranges for the different candidate solar-sail material we can find for which energies for protons and α-particles the solar sail is transparent**.** For example, $3.70\times10^{-6}$g/cm$^2$ is the range for 20 nm beryllium foil and we can conclude that the beryllium sail wall is transparent to protons and α-particles with energy higher than 10 keV.

As a result of ionization processes produced mostly by UV radiation and also by the solar electrons, protons and α-particles the solar sail is gaining a positive electric charge. As it is shown in Ref. 7 the solar electrons recombination with sail material positive ions can not neutralize the solar sail. The relative rates of the ionization and recombination processes determine the establishment of the positive charge on the solar sail, which constantly increase during the perihelion pass.

The ionized sail may have one advantage for both classes of solar sails related to the electrical propulsion by the solar wind, in other words, for sailing in the solar wind, as well as one disadvantage related to the electrostatic pressure in the case of a hollow-body solar sail. In Ref. 14 proposed an electrical sail and analyzed an interplanetary spacecraft consisting of a positively charged grid that electrically reflects incident solar wind protons and an onboard device that maintains the grid's potential. As it is shown above in our case as a result of UV ionization the solar sail naturally gains the positive potential which will deflect the proton component of the solar wind and therefore, extract momentum from the solar wind plasma. However, the proton component of the solar wind induced acceleration is considerably less that the radiation-pressure acceleration [7] and be a minor perturbation rather than a large effect. On the other hand this minor perturbation should be taken into consideration for the calculation of the trajectory for an interstellar sailcraft.

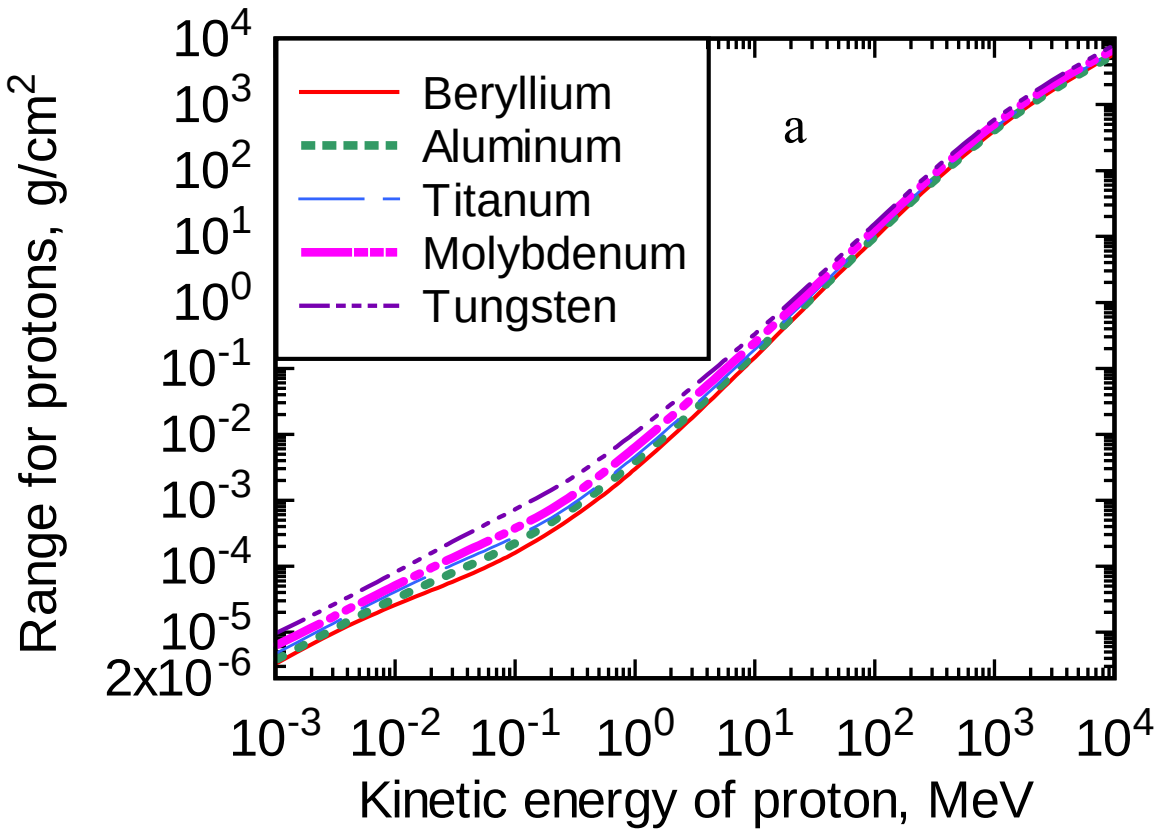

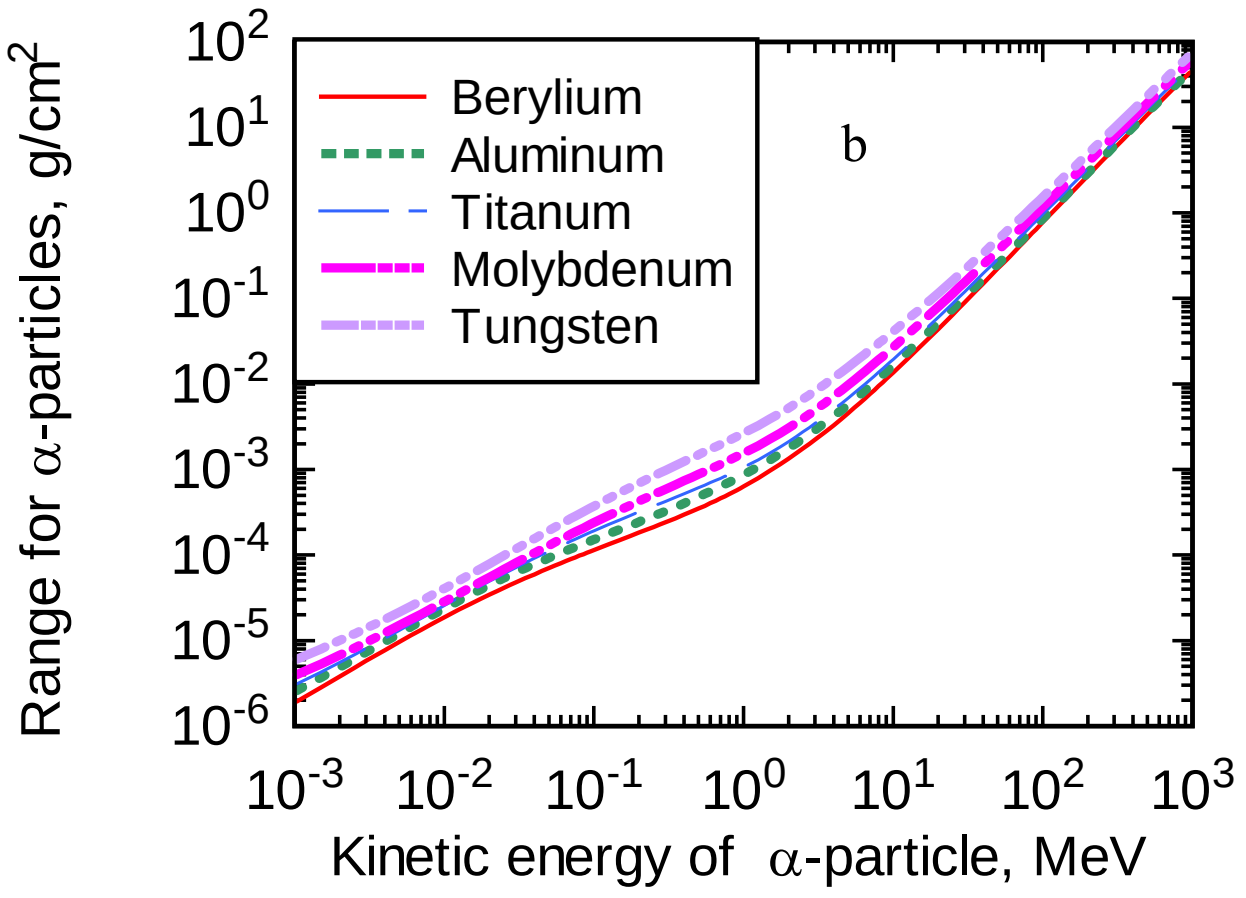

Fig. 5. The energy dependence of the stopping range for the protons (a) and α-particles (b) for a different candidate solar-sail.

The ionized sail surface has a surface charge distribution which is producing the electric field $E$. The electric field of the solar sail very near the surface can be considered as an electric field produced by an infinite plane. The electric field inside of the hollow-body solar sail according to Gauss' law is zero. In the presence of an electric field, a surface charge on the sail will experience a force per unit area which is an outward electrostatic pressure which can also be written as

$$P = \frac{\varepsilon_0}{2} E^2 , \qquad (6)$$

where $\varepsilon_0$ is the permittivity of free space. Electrostatic pressure acting on the surface of the sail always acts outward. In Ref. 7 we found that for beryllium hollow-body solar sail $P = 59.6$ MPa. It is about the same as the tensile strength of beryllium at temperature 800 $^0$C. The tensile strength depends on the temperature. When temperature increases the tensile strength of materials usually decreases. Hence, the electrostatic pressure can exceed the solar sail material's tensile strength, and therefore the surface of the hollow-body solar sail may fragment. This effect can be visualized by charging up soap bubbles or spherical rubber balloon: the additional electrostatic pressure eventually causes them to burst.

## 6. Diffusion of inflation gas in hollow-body sail

Different processes occur in the inflation fill gas within the solar sail. A fraction of fill gas molecules striking the inner surface of the sail are directly scattered back. The number of molecules promptly returned back by this scattering process depends on their incident energy, therefore on the temperature of the gas, as well as the type of material of the sail walls. The fill gas molecules that are stopped in the sails walls have several possible fates. Previously existing defects, defects caused by the energetic solar electrons and protons may trap hydrogen. These traps are defined by the maximum range of the incident hydrogen. The traps eventually become saturated, and the untrapped gas diffuses, going deeper into the sail walls' material or in to space. Below, we consider the atomic diffusion - the process whereby the random thermally-activated gas atoms and ions result in net transport through the sail foil. In other words, gas molecules, atoms and ions inside the hollow-body solar sail can diffuse through the wall of the sail and escape, resulting in the sail slowly deflating. The diffusion equation contains two gas-metal interaction parameters, the diffusivity $D$, which describes the gas transport in a concentration gradient and the heat transport $Q$, which describes the gas transport in a temperature gradient. However, because the thickness of the solar sail wall is about a few hundred atomic layers the temperature gradient is negligibly small and in a very good approximation we can consider that $\nabla T = 0$ and neglect the term describing the heat transport through the walls in the diffusion equation. There is a concentration gradient in the sail wall, because the sail is filled initially with gas, and there is no gas on the outside. Therefore, the rate of transport of hydrogen is governed by the diffusivity and the concentration gradient. For simplicity let us consider the diffusion flux perpendicular to the wall and Fick's first law for the a diffusion flux can be written in the form

$$j = D_0 e^{-\frac{E_D}{kT}} \frac{dC}{dx}. \qquad (7)$$

where $D_0$ is the diffusion coefficient or diffusivity, $C$ is the concentration which are measured in mol/m$^3$ and m$^2$/s, respectively, $T$ is absolute temperature of the hydrogen gas, $E_D$ is the activation energy for the diffusion and $k = 8.63\times10^{-5}$ eV/K is the Boltzmann's constant.

The fill gas is in thermal equilibrium with the solar sail walls. Therefore, the gas temperature will be equal to the temperature of the solar sail foil. In the worst case scenario this temperature should be close to the sail material melting temperature.

The thickness of the hollow-body solar sail for the Sun-facing surface and the anti-solar sail surface may differ depending on its design [4, 5]. However, the thickness of the sail material will not exceed a few hundred atomic layers. For such small thicknesses in good approximation we can replace the derivative of the concentration by the concentration difference over thickness $d$, assuming that concentration of the hydrogen atoms deceases uniformly and consider that the outside concentration of the hydrogen for the diffusion flux is zero for both the Sun-facing sail surface, and the anti-Sun solar sail surface, we can rewrite Eq. (7) in the following form

$$j = D_0 e^{-\frac{E_D}{kT}} \frac{C_{in}}{d}, \qquad (8)$$

where $C_{in}$ is the concentration of gas into the hollow-body solar sail. Equation (8) shows the dependence of the gas diffusion flux on the temperature and thickness of the wall for the solar sail. The total molecular flux rate of the gas through the all sail area is

$$\frac{dM}{dt} = \oiint \vec{j}\vec{n}dS, \qquad (9)$$

where $\vec{n}$ is the normal to the sail surface $S$. Using Eq. (8) and considering that the hollow-body sail has the shape of a disk of radius of $R$ and height $h$, we can calculate the total gas mass flux rate given by Eq. (9). Let's also mention that using Eqs. (8) and (9) we can determine the dependence of the total molecular flux rate of the gas on the temperature of the solar sail material. As an example, we performed calculations for the hollow-body beryllium sail considered in Ref. 5 for the thickness of the beryllium foil of 40 nm which follows from our calculation in Sec. 4. Fig. 6 present the results of our calculations for the dependence of the flux rate of the hydrogen on the temperature for two configurations of the sailcraft. We consider the different values of the diffusion activation energy $E_D$ and constant $D_0$ The experimental value of the activation energy $E_D$ and constant $D_0$ for the diffusion of beryllium depends on the purity of the beryllium material used in the experiments and varies from 0.15 eV to 0.36 eV and from $9\times10^{-12}$ m$^2$/s to $6.7\times10^{-9}$ m$^2$/s, respectively [15-18]. We calculated the flux rate for different available sets of the diffusion activation energy and diffusion constant and presented in Fig.1 by thin curves for configuration A and by thick curves for configuration B. A general overview of these results shows that the diffusion flux rate strongly depends on the temperature. When temperature increases from 400 K to 1400 K, the flux rate increases by a factor of $10^4$ and $10^3$ for the solar sail configuration A and B, respectively, when in calculations used the diffusion activation energy and diffusion constant from Ref. 16. For these parameters from Refs. 15 and 16, the hydrogen flux rate is changing by a factor of $10^2$ and almost by a factor of 10 for the solar sail configuration A and B, respectively. Over temperature 500 – 1400 K, hydrogen flux is changing more than three orders of magnitude for the diffusion activation energy and diffusion constant reported in Ref. 17 for the configuration A, as well as for the configuration B. Hence, we can conclude that the results also demonstrated the strong dependence of the flux rate on the values of the diffusion activation energy and diffusion constant.

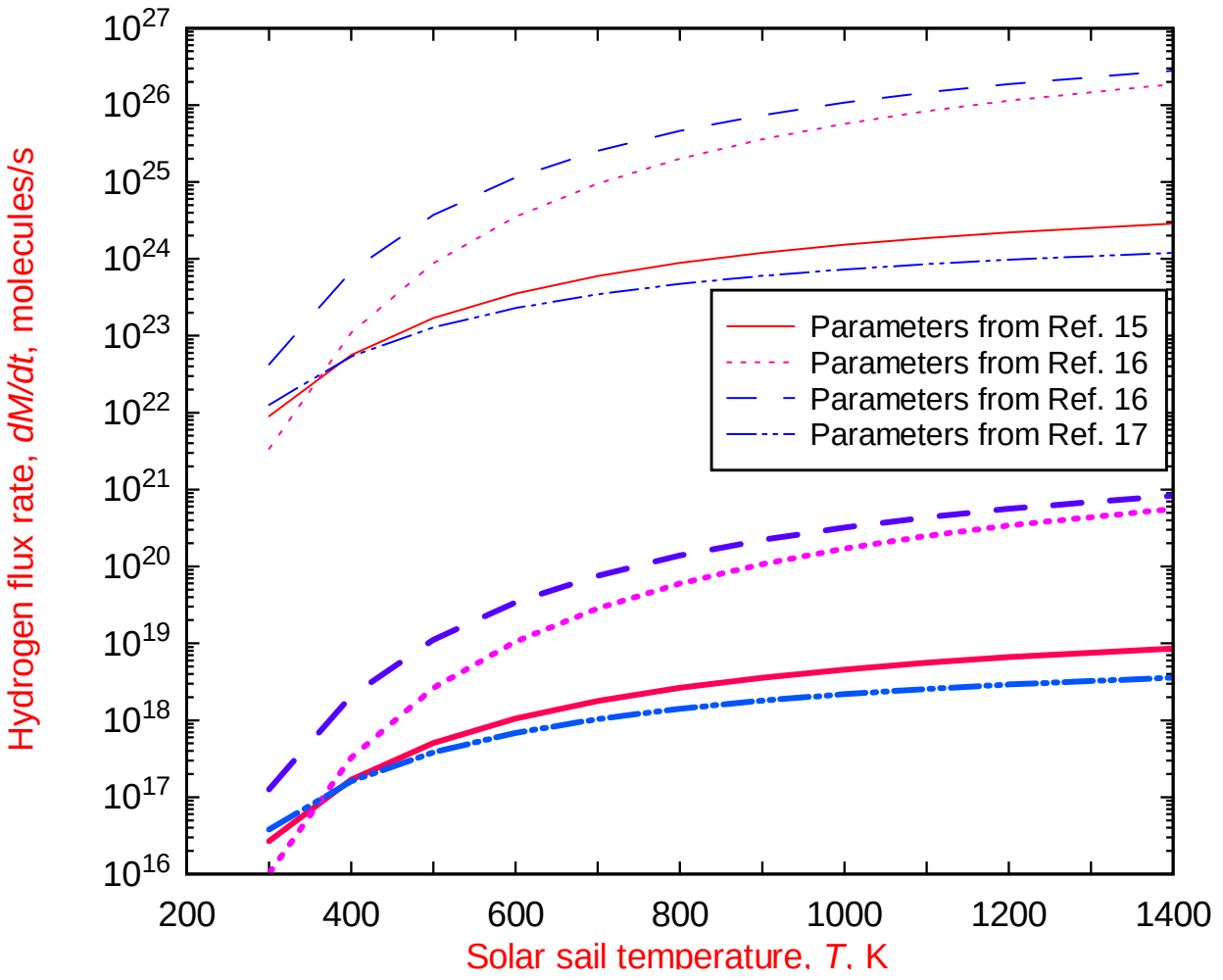

Fig. 6. Temperature dependence of hydrogen flux rate for beryllium solar sail configuration A (thin curves) and configuration B (thick curves) Results of calculations are given for the different parameters of diffusion activation $E_D$ and diffusion constant $D_0$.

Our approach allows us to estimate how much of the inflation fill gas should be carried by the hollow-body solar sail for the acceleration phase in the near-Sun space-environment. To illustrate that let us consider the configuration A and B of the beryllium hollow-body solar sail. The total amount of hydrogen molecules in the spacecraft configuration A and B is $9.2 \times 10^{30}$ and $5.0 \times 10^{22}$, respectively [5]. The acceleration time of the sailcraft nearby the Sun is about 7000 s - 8000 s. Using these values we can estimate the performance time for the solar sail. From Fig.6 it is easy to see that negligible amount of the hydrogen filled gas will diffuse through the walls of the beryllium sail for configuration A for the temperature regime up to 1400 K, when we used in our calculations the diffusion activation energy and diffusion constant from Refs. 15-17. However the same analysis of Fig. 6 shows that the sail will be deflated completely for configuration B when the temperature of the sail exceeds 400 K - 450 K for the sets of parameters from Ref. 16. For the diffusion activation energy and diffusion constant from Refs. 15 and 17, the solar sail for the configuration B will deflate completely when the temperature of the sail walls exceeds about 1300 K.

## 7. Conclusions

To find minimal required thickness of the solar sail for an acceleration of the sail craft in the near-Sun space region the following important factors should be considered: the existence of a wide range of solar electromagnetic radiation frequencies, dependence of the electrical conductivity and dielectric function of a sail material on the frequency and the temperature dependence of electrical conductivity of a sail material.

In the present work we introduced the results of the interactions of solar electromagnetic radiation and solar electrons, protons and α-particles with the different candidate solar-sail materials. Some calculations are emphasized on the beryllium hollow body sail with the hydrogen fill gas. Our calculations were performed using most conservative assumptions with the highest values for the available cross sections for the interaction of solar photons, electrons, protons α-particles with solar sail material atoms and nuclei. Under these assumptions we found that mostly the solar deep UV radiations will ionize solar sail foil.

Moreover, the accumulation of the electric charge on the solar sail due to UV ionization will constantly increase for the duration of the operation of the solar sail and produces the electrostatic electric field and therefore, a positive electric potential. This electric potential will deflect the proton component of the solar wind and therefore, extract momentum from the solar wind plasma and accelerate the solar sail. This acceleration is a minor perturbation with respect to the solar radiation acceleration and should be taken into consideration for the calculation of the trajectory for the sailcraft for a long term interstellar travel. The electric field is excluded from the inside of the hollow-body solar sail, but not from the outside, giving rise to a net outward force producing the electrostatic pressure. The outward electrostatic pressure on the hollow-body sail surface can exceed the tensile strength of the solar sail material, and as a result, the surface of the hollow-body solar sail may burst.

Our consideration of the worst scenario for the diffusion of hydrogen through the solar sail beryllium walls show that the diffusion flux rate strongly depends on the temperature. The results also demonstrated the strong dependence of the flux rate on the values of the diffusion activation energy and diffusion constant.

Thus, in our microscopic approach to obtain a high performance solar sail as a first step, when the perihelion of an initially elliptical transfer orbit to a parabolic or hyperbolic solar orbit is found, is to select among materials with the same optical properties, the material with low density and find the thickness of the foil of this material based on the existing solar sail equilibrium temperature at the perihelion. Then for that thickness of the solar sail we can determine the ionization degree of the solar sail material and find the electric charge accumulated during the perihelion pass. For a hollow-body solar sail for the given sail thickness we should determine the amount of the inflation gas that will not diffuse through solar sail wall and guarantee the perihelion pass. This approach can be applied to a wide variety of materials, and sail configurations utilizing metallic thin foils in the near-Sun space environment.